# What Can We Estimate from Fatality and Infectious Case Data using the Susceptible-Infected-Removed (SIR) model? A case Study of Covid-19 Pandemic

**Semra Ahmetolan[1], Ayse Humeyra Bilge[2], Ali Demirci[1*], Ayse Peker-Dobie[1,2], Onder Ergonul[3]**

[1]Department of Mathematics, Faculty of Science and Letters, Istanbul Technical University, 34469, Istanbul, Turkey
[2] Department of Industrial Engineering, Faculty of Engineering and Natural Sciences, Kadir Has University, 34083, Istanbul, Turkey
[3]Infectious Diseases and Clinical Microbiology Department, School of Medicine, Koc University, Istanbul, Turkey

**\* Correspondence:**
Ali Demirci
demircial@itu.edu.tr

## Abstract

The rapidly spreading Covid-19 that affected almost all countries, was first reported at the end of 2019. As a consequence of its highly infectious nature, countries all over the world have imposed extremely strict measures to control its spread. Since the earliest stages of this major pandemic, academics have done a huge amount of research in order to understand the disease, develop medication, vaccines and tests, and model its spread. Among these studies, a great deal of effort has been invested in the estimation of epidemic parameters in the early stage, for the countries affected by Covid-19, hence to predict the course of the epidemic but the variability of the controls over the course of the epidemic complicated the modeling processes.

In this article, the determination of the basic reproduction number, the mean duration of the infectious period, the estimation of the timing of the peak of the epidemic wave is discussed using early phase data. Daily case reports and daily fatalities for China, South Korea, France, Germany, Italy, Spain, Iran, Turkey, the United Kingdom and the United States over the period January 22, 2020 - April 18, 2020 are evaluated using the Susceptible-Infected-Removed (SIR) model. For each country, the SIR models fitting cumulative infective case data within 5% error are analysed. It is observed that the basic reproduction number and the mean duration of the infectious period can be estimated only in cases where the spread of the epidemic is over (for China and South Korea in the present case). Nevertheless, it is shown that the timing of the maximum and timings of the inflection points of the proportion of infected individuals can be robustly estimated from the normalized data. The validation of the estimates by comparing the predictions with actual data has shown that the predictions were realised for all countries except USA, as long as lock-down measures were retained.

# 1    INTRODUCTION

The Coronavirus disease 2019 (Covid-19) caused by severe acute respiratory syndrome coronavirus 2 (SARS-CoV-2) is a highly contagious disease affecting huge numbers of people all over the world. The earliest case was identified in China in December 2019. After the first diagnosis, the disease has spread very quickly to other countries, in spite of efforts to slow and stop the transmission of COVID-19, such as self-isolation, quarantine, social distancing, contact tracing, and travel limitations. As a result of its rapid spread and very high infection rates, the World Health Organization (WHO) declared Covid-19 a pandemic in March 2020 [1].

As of April 2020, even though the pandemic has passed its early stage and there are 90% fewer cases in China as a consequence of successful containment measures, the disease is rapidly expanding in Europe, America, Asia, Middle East and Africa. Despite the application of travel restrictions by many countries, there have been no substantial delays in the arrival of the pandemic in non-affected areas, as in the case of the H1N1 epidemic in 2009 [2].

A great deal of effort has been invested in the estimation of epidemic parameters of Covid-19 in the early stage for China and some other countries [3], [4], [5], [6], [7], [8], [9], [10], [11], [12], [13]. In [3], the authors analysed the temporal dynamics of the disease in China, Italy and France in the period between 22nd of January and 15th of March 2020. In [4], the potential for sustained human-to-human transmission to occur in locations outside Wuhan is assessed based on the estimations of how transmission in Wuhan varied between December, 2019, and February, 2020. The difficulties related to the accurate predictions of the pandemic is discussed in [5]. In [6], the authors used phenomenological models that were developed for previous outbreaks to generate and assess short-term forecasts of the cumulative number of confirmed reported cases in Hubei province and for the overall trajectory in China [7]. Epidemic analysis of the disease in Italy is presented in [8] by means of dynamical modelling [9]. Forecasting Covid-19 is investigated in [10] by using a simple iteration method that needs only the daily values of confirmed cases as input. A cumulative distribution function (CDF) and its first derivative are used to predict how the pandemic will evolve in [11]. In [12], the authors proposed a segment Poisson model for the estimation. In [13], a meta-population model of disease transmission in England and Wales was adapted to predict the timing of the peak of the epidemic. In addition, it was shown that the change in the epidemic behaviour of various countries can be traced by the use of data driven systems [14].

One of the common features of these works is the existence of variations in these parameter estimations. In the present work, the determination of the following parameters is discussed:

1) The Basic Reproduction Number $\mathfrak{R}_0$,
2) The mean duration of the infectious period T,
3) The time $t_m$ (days) at which the number of infectious cases reaches its maximum, i.e, the first derivative of I(t) is zero,
4) The time $t_a$ (days) at which the rate of increase in the number of infectious cases reaches its maximum, i.e., the time at which the second derivative of I(t) is zero and the first derivative is positive,
5) The time $t_b$ (days) at which the rate of decrease in the number of infectious cases reaches its maximum, i.e., the time at which the second derivative of I(t) is zero and the first derivative is negative.

By employing the Susceptible-Infected-Removed (SIR) model, we show that the quantity that can be most robustly estimated from normalized data, is the timing of the maximum and timings of the inflection points of the proportion of infected individuals. These values correspond to the peak of the epidemic and to the highest rates of increase and the highest rates of decrease in the number of infected individuals. The stability of the estimations is discussed by comparing predictions based on data with long time spans.

## 2  DATA AND METHODS

Publicly accessible data that have been released by the state offices of each country are used for the analysis. The data set of each country is collected according to published official reports and available at the website http://www.worldometers.info/coronavirus/ (last access: 27 April 2020). Updated data are also available at the website http://epikhas.khas.edu.tr/. Data used for the analysis covers the period January 22-April 18, 2020 and in the following, "Day 1" corresponds to January 22, 2020. The analysis uses Susceptible-Infected-Removed (SIR) model [15] and solutions are obtained by numerical methods. Updated data covering the period 19 April-1 July is used to assess the performance of the model.

### 2.1  SIR model

The Susceptible-Infected-Removed (SIR) model is a system of ordinary differential equations modelling the spread of epidemics in a closed population, under the assumption of permanent immunity and homogeneous mixing [15]. These equations are

$$S' = -\beta\, S\, I, \quad I' = \beta\, S\, I - \eta\, I, \quad R' = \eta\, I. \tag{1}$$

Since the right hand sides of these equations add up to zero, the sum S+I+R is a constant that is equal to the total number of individuals in the population. Thus by normalizing, we may assume that S, I and R are proportions of individuals in respective groups. Since the Covid-19 infection has an incubation period, the right model to use is the SEIR system. But, in previous work [16] it was shown that the parameters of the SEIR model cannot be determined from the time evolution of the normalized curve of removed individuals. Thus the SEIR model should not be used in the absence of additional information that might be obtained by clinical studies. In the present work, since we assume no clinical information we will use the SIR model, with necessary modifications for the interpretation of the results, as indicated in [16].

### 2.2  Relation between the basic reproduction number and the total number of removed individuals

The ratio $\beta/\eta$, called the Basic Reproduction Number and denoted as $\Re_0$, is the key parameter in both the SIR and SEIR models. This number is related to the growth rate of the number of infected individuals in a fully susceptible population and determines the final value of R denoted by $R_f$ that is the proportion of individuals that will be affected by the disease. This proportion includes individuals who gain immunity without showing symptoms, those who are treated, as well as disease-related fatalities. The reciprocal of the parameter $\eta$, $T = 1/\eta$ is considered as a representative of the mean infectious period.

The relation between $\Re_0$ and $R_f$ is determined as follows. Note that R(t) is a monotonically increasing function, and hence it can be used as an independent variable, instead of t. The derivative of S with respect to R is given by

$$dS/dR = S'/R' = -\beta/\eta \; S = -\Re_0 \; S. \qquad (2)$$

Assuming initial conditions $S \rightarrow 1$ and $R \rightarrow 0$ as t approaches negative infinity, on can integrate and obtain $S(t) = e^{-\Re_0 R(t)}$. Then, as t approaches positive infinity, since $I \rightarrow 0$, S+R=1 yields

$$R_f + e^{-\Re_0 \; R_f} = 1. \qquad (3)$$

$\Re_0$ can be solved from this equation as a function of $R_f$, and their relation is displayed on Figure 1.

The graph of $R_f$ versus $\Re_0$ is shown on Figure 1, together with the ranges of $\Re_0$ for well-known diseases. It can be seen that for $\Re_0 > 2.5$, $R_f$ is greater than 90%. The figure also shows that the increase in $R_f$ with respect to $\Re_0$ is very slow for $\Re_0 > 3$. It is generally accepted that the $\Re_0$ for Covid-19 is greater than 3 despite all containment measures [17], [18], [19]. Thus, unless vaccination is applied, one would expect that at least 95% of the population would be affected by the disease. In addition, the knowledge of its precise value would have little effect on the planning of healthcare measures. It should also be kept in mind that containment measures provide a temporary control of the spread of the epidemic, just to the point of reducing the burden of the epidemic to a manageable size.

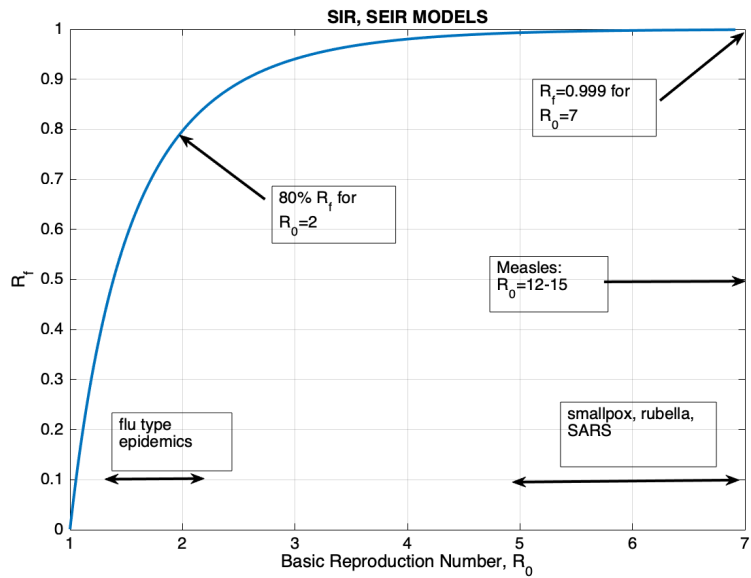

Figure 1. The graph of $\mathbf{R_f}$, the final proportion of individuals that would be affected by the disease, versus the Basic Reproduction Number $\boldsymbol{\Re_0}$.

According to the Centers for Disease Control and Prevention (CDC), at the time we completed the data collection phase of our research, it was still unknown when viral shedding begins or how long it lasts for, and nor is the period of COVID-19's infectiousness known. Like infections with MERS-CoV and SARS-CoV, SARS-CoV-2 RNA may be detectable in the upper or lower respiratory tract for weeks after illness onset, though the presence of viral RNA is no guarantee of the presence of the infectious virus. It has been reported that the virus was found without any symptoms being shown (asymptomatic infections) or before symptoms developed (pre-symptomatic infections) with SARS-Cov-2, though the role they may play in transmission remains unknown. According to prior studies, the incubation period of SARS-CoV-2, like other coronaviruses, may last for 2-14 days [20].

To illustrate an example for an SIR model, $\Re_0$, T and R(0) are chosen as 3, 10 and $10^{-3}$, respectively and the related graphs are given on Figure 2.

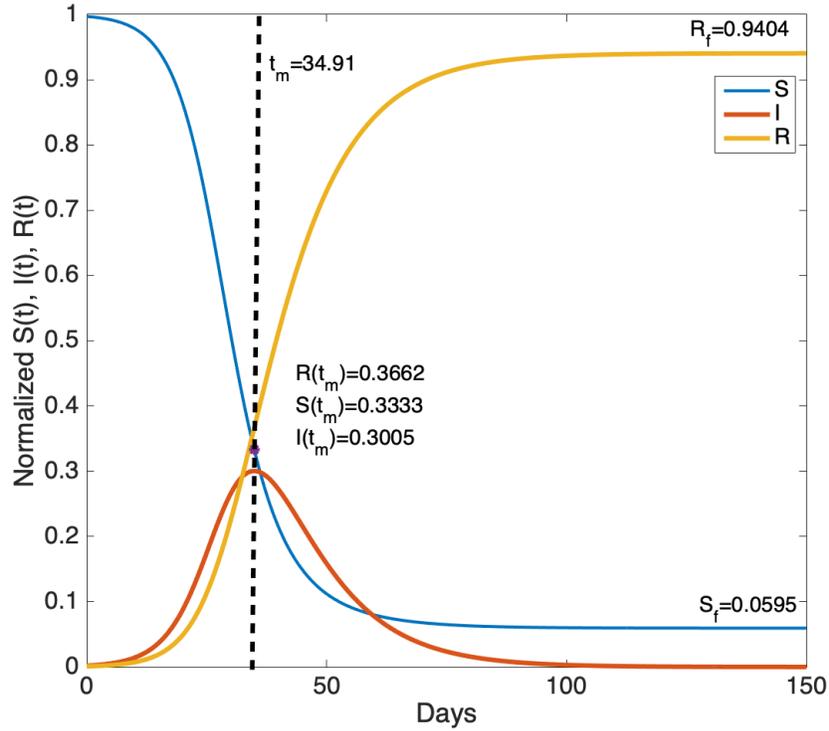

Figure 2: The time evolution of S(t), I(t) and R(t) for $\Re_0$=3, T=10 days and R(0)= $10^{-3}$.

From Figure 2, it can be seen that for parameter values $\Re_0$=3, T=10 day, the duration of the epidemic is about 100 days. The peak of the epidemic occurs approximately at day 35. Note that the derivative of I(t) vanishes at time $t_m$ when S($t_m$)= 1/ $\Re_0$. In this example, S($t_m$)= 0.3333, I($t_m$)=0.3005 and R($t_m$) =0.3662. The final values of S(t) and R(t) are $S_f$=0.0595 and $R_f$=0.9404 at the end of the epidemic.

### 2.3 Representative data for the proportion of removed individuals

It is in general accepted that the number of fatalities represents the number of removed individuals and the number of confirmed cases represents the number of infected individuals. In the initial phase of the epidemic, little information was available on the proportionality constants, but as long as they don't change in time, one can work with the normalized case reports and normalized fatalities and look for the determination of the epidemic parameters from the shape of these normalized curves. In Section 4, it will be shown that for the Covid-19 data, total cases would be a better representative of the number of removed individuals.

## 3    OVERVIEW OF DATA

According to the SIR model, given by the equations in (1), the rate of change of the number of removed individuals is proportional to the number of infectious cases. In terms of observations, this corresponds to the fact that the ratio of, for example, daily fatalities to daily infectious cases should be constant. In the literature on the analysis of historical epidemics, fatality reports are usually the only available data, hence models are necessarily based on the assumption that cumulative fatalities represent cumulative number of removed individuals. For the Covid-19 pandemic, as daily fatality and infectious case reports are available, further evaluation of the representation of R(t) in terms of fatality data is presented. Daily infections and total fatalities are displayed on Figure 3, for all countries.

### 3.1    Time evolution of daily infections and total fatalities

Normalized daily infectious cases and total fatalities are shown on Figure 3.

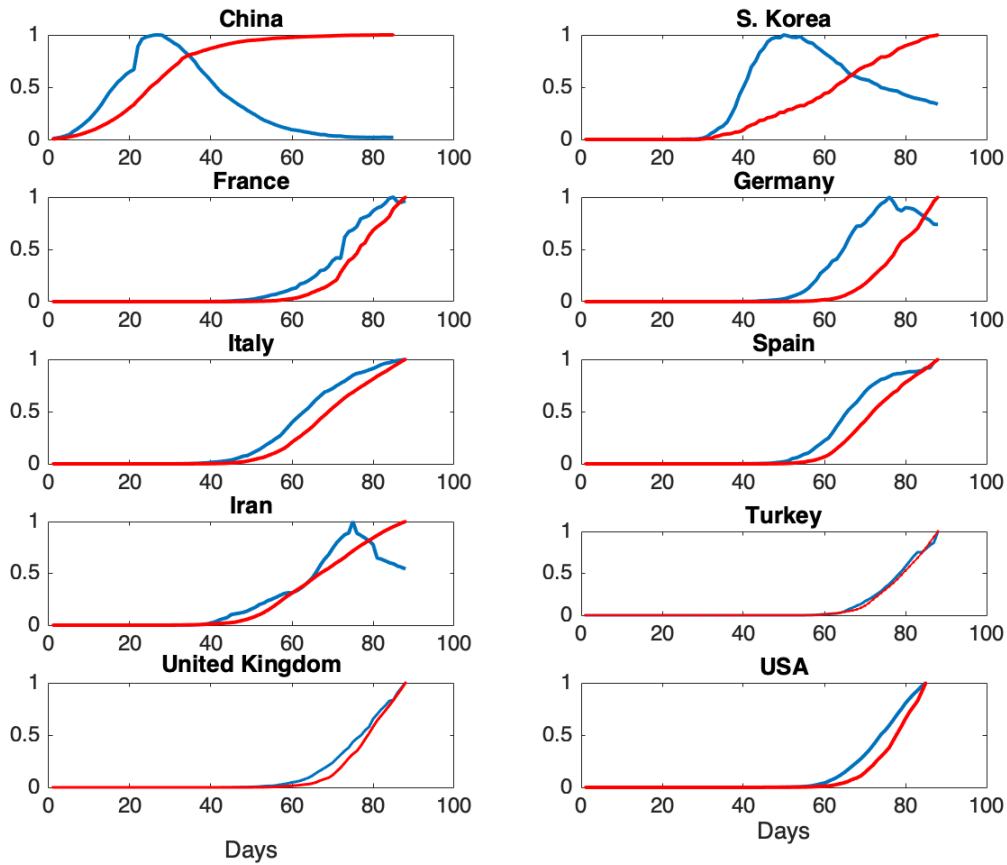

Figure 3.  Normalized daily case reports (Blue) and normalized total fatalities (Red) for each country. The horizontal axis represents days starting from January 22 and vertical axis represents ratios.

From Figure 3, it can be seen that the epidemic cycle has been completed in China over the course of about 70 days. The jump in total fatalities is due to a change in the reporting scheme. As our analysis is based on total infectious cases, this change has no effect on the models. For South Korea, the epidemic is in a state of slow decrease at the end of about 60 days, but the rate of infections is still high. This qualitative behaviour is an indication of the fact that $\Re_0$ for South Korea is expected to be much higher than the one for China [7, 21]. For France, Germany and Iran, the epidemic is in the decline phase. For the rest of the countries, further analysis is needed in order to assess epidemic phase.

## 4    ESTIMATION OF THE SIR MODEL PARAMETERS

As noted above, the knowledge of $\Re_0$ determines the total proportion of individuals that would be affected, $R_f$. Furthermore, the peak of I(t) occurs at the time $t_m$, at which the proportion of susceptible individual falls to the value $1/\Re_0$. This information is useful for the determination of the proportion of people that have to be vaccinated in order to drag the proportion of susceptible individuals below this threshold. The Basic Reproduction Number is "defined" as the number of new infections per unit time in a fully susceptible population. Thus, it is a quantity that might be measured by direct on-site observations. On the other hand, the knowledge of $\Re_0$ by itself does not give any information on the timing of the progress of the epidemic.

It will be seen that $\Re_0$ and T can be estimated only for China where the spread of the epidemic is over. For other countries, $\Re_0$ and T cannot be estimated from the normalized data, but the timings of the key events, $t_m$, $t_a$ and $t_b$ can be determined quite reliably.

### 4.1    Methods for estimating the parameters $\Re_0$, $T$, $t_m$, $t_a$, and $t_b$

These parameters are determined by a "brute force" approach: The models are run for a broad range of parameters. Then the difference between data and the model is compared by using various norms. Finally, the models that match data within 5% are selected. If the scatter plot of the errors versus the parameter to be estimated has a sharp minimum, it is concluded that the corresponding parameter can be determined from the shape of the normalized data.

The parameter ranges for the SIR model are

$$\mathbf{1.5 < \Re_0 < 10}, \qquad \mathbf{2 < T < 30}, \tag{4}$$

and the initial values are chosen as

$$\boldsymbol{R_{ini} = 10^{-k}}, \qquad \boldsymbol{S_{ini} = e^{-(\beta/\eta)R_{ini}}}, \qquad \boldsymbol{I_{ini} = 1 - S_{ini} - R_{ini}}, \tag{5}$$

where $1 < k < 10$. For South Korea, these parameter ranges are extended appropriately.

### 4.2    Selection of representative Data for R(t)

In the SIR model, since $\boldsymbol{R' = \eta\,I}$; that is, the rate of change in the number of removed individuals is proportional to the number of infected individuals, it is expected that the cumulative cases are proportional to cumulative fatalities. Thus, the SIR model predicts the simultaneity of the daily fatalities and daily infections. The verification of this fact requires the availability of data both for infections and for fatalities. The data for the 2009 H1N1 epidemic collected at certain major hospitals [22] is valuable in the sense of reflecting information on both infections and fatalities. The peculiarity of this data is a shift of about 8 days between total infections and total fatalities, the peak of infections occurring 8 days prior to the peak of fatalities. This time shift was explained by a multi-stage SIR model [23].

Cumulative cases and cumulative fatalities for Covid-19 do not show such a clear time shift. On the contrary, in China and Korea, fatalities increase faster than infections. In Germany, there is a slight lead for infections, while for other countries the two curves more or less coincide. The lead of fatalities over infections that is observed in China and in Korea is an unexpected fact, which is possibly due to the irregularities in the statistics, in medical treatment practices, etc. We should also note that the progression of the Covid-19 epidemic is unique in the sense that new treatment methods are applied during the initial phase in China and these methods have been applied in other countries.

For China, several programs were run, first by fitting the predicted R(t) to the total fatality data, then to the cumulative infectious case data. In the first case, about 700 models fitting cumulative fatalities within 5% error and about 3000 models that fit cumulative infections within 5% error are found. Furthermore, in the latter case, the minima for the quantities that were aimed to be determined were much sharper. For South Korea, as it will be explained later, the model matching was not successful. For other countries, as the difference between total infections and total fatalities was negligible, total infections are used as a representative of R(t) of the SIR model.

Our main result is that it is not possible to determine the Basic Reproduction Number and the mean duration of the infectious period from the shape of the *normalized* data (unless there are reasonable estimates for either of these parameters). In order to make a reliable determination of the parameters $\mathfrak{R}_0$ and T by using the early stage data, a certain period of time has to pass. This period is approximately 70 days for 2009 A(H1N1) epidemic [23]. However, this period for Covid-19 is still uncertain. This is possibly the reason why the parameters for countries other than China and South Korea can not be established. On the other hand, the timings of the peak of the infectious cases, the peak of the rate of increase and the rate of decrease of the infectious cases can be determined more precisely from the shape of the normalized data.

### 4.3 Simulations for SIR models with $\mathfrak{R}_0$/T =constant

The 'best' estimations of the parameters $\mathfrak{R}_0$ and T lie on a curve that is nearly linear when a SIR model is used to fit the data of an epidemic. This fact has been observed in previous work [24], in the study of the H1N1 epidemic and it was explained by the fact that the duration of the epidemic pulse (appropriately defined in terms of a fraction of the peak of infections) was nearly invariant for values of $\mathfrak{R}_0$ and T, with $\mathfrak{R}_0$/T constant.

In order to visualize this situation, the solutions of this system of differential equations of the SIR model (1) for parameter range 3< $\mathfrak{R}_0$ <20, and $\beta = \mathfrak{R}_0$/T = 1/5 are obtained. The graphs of normalized solutions (after an appropriate time shift) are given in Figure 4.

### 5 RESULTS FOR EACH COUNTRY

The scatter plots of the mean infectious period T versus $\mathfrak{R}_0$, and the scatter plots of the modelling error versus the parameters are presented in Figures 5-9 where $I_t$ and $I_{tt}$ represent the values of the first and the second derivatives of I(t) at the last day of the data April 18, 2020, respectively. The error stands for the relative error between the normalized R(t) of the model and normalized total infectious cases, in the $L_2$ norm.

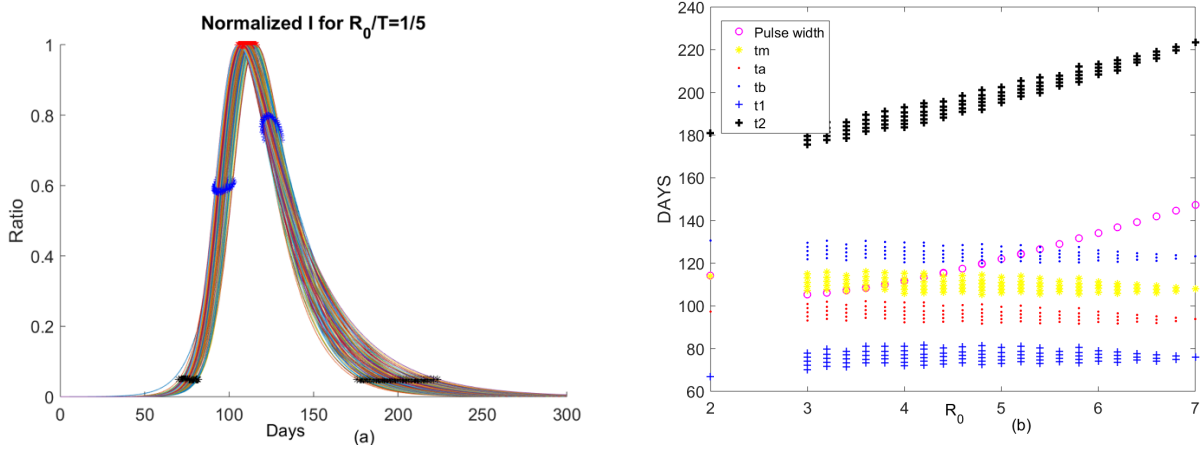

Figure 4 (a) Normalized values of I(t) for 3< $\mathfrak{R}_0$ <20, and $\boldsymbol{\beta} = \mathfrak{R}_0/T = 1/5$, together with the inflection points ($t_a$, $t_b$), the peak point ($t_m$) and the timing of the initial ($t_1$) and final ($t_2$) points when the 5% of maximum value I(t) epidemic, (b) Dependency of $t_1$, $t_a$, $t_m$, $t_b$, $t_2$ on $\mathfrak{R}_0$.

### 5.1 *Scatter plot of the mean duration of the infectious period versus the basic reproduction number*

In Figures 5-9, the first graph, in the upper left of the panel is the scatter plot of the mean duration of the infectious period, T, with respect to the basic reproduction number $\mathfrak{R}_0$, for models that fit data within 5% error in the norm described above. For all countries, the "best" parameters lie on a curve, instead of being agglomerated around a mean. This indicates that although the SIR model fitting normalized data is unique, the parameters $\mathfrak{R}_0$ and T cannot be determined precisely from normalized data. The colors blue, red and yellow in Figures 5-9 represent the results according to whether the last day of the analysis, $t_f$, is 78, 83 and 88, respectively.

### 5.1.1 Scatter plot of the modelling error versus the basic reproduction number and versus the mean duration of the infectious period

In Figures 5-9, the second (first row, right panel) and the third (second row, left panel) graphs display the scatter plot of the modelling error with respect to $\mathfrak{R}_0$ and T respectively. For China, there are well defined minima in the modelling errors at nearly $\mathfrak{R}_0$=3 and T=9. For South Korea, the minima of the error in $\mathfrak{R}_0$ seems to be located beyond $\mathfrak{R}_0$=8, and the minimal error in T corresponds to T=25 approximately. These parameter values are not in the ranges reported in the literature. Data for South Korea shows different characteristics, that might be due to the strategy of extensive testing and filiation, as opposed to lock-down measures. An indication of extensive testing policy is the fact that approximately 27.4 percent of confirmed coronavirus patients in South Korea were in their 20s, showing that asymptomatic cases are also included in the statistics. For all of the remaining countries, the ranges of $\mathfrak{R}_0$ and T corresponding minimal modelling errors are too large to attempt any reasonable estimation for these parameters. If either $\mathfrak{R}_0$ or T is estimated by using alternative methods (medical observations etc), it would be possible to obtain better estimates and improve the model by bootstrapping.

### 5.1.2 Timing of the peak of the maximum for I(t)

The fourth (second row, right panel) graph in Figures 5-9 shows the scatter plot of the modelling error versus $t_m$, the timing of the peak of the number of infections. For all of the countries analysed, this parameter can be estimated quite sharply. In order to study the reliability of this estimation, the model matching process is repeated for $t_f$ =78, 83 and 88.

### 5.1.3 Timing of the inflection points of I(t)

The ratio of infected individuals I(t) has two inflection points. The first inflection point ($t_a$) is located at the left of the maximum ($t_m$) whereas the second one ($t_b$) is located at the right of $t_m$. $t_a$ and $t_b$ correspond to the highest rate of increase and decease in I(t), respectively. In Figures 5-9, the right and left panels of the third row display scatter plot of the error in these quantities. Their variation with respect to $t_f$ is also investigated.

### 5.1.4 Final values of the first and second derivatives of I(t)

The values of the first and second derivatives at $t_f$ are shown on the fourth row, left and right panels, respectively. If the first derivative is positive (negative), the I(t) is in the rising (falling) phase, while if the second derivative is positive (negative) the curve is concave up (down).

The epidemic phases which are shown in Figure 10, are categorized by the sign of the first and the second derivatives of I(t) as follows

1. Phase I: slow increase $\left(\frac{dI}{dt} > 0, \frac{d^2 I}{dt^2} > 0\right)$
2. Phase II: fast increase $\left(\frac{dI}{dt} > 0, \frac{d^2 I}{dt^2} < 0\right)$
3. Phase III: fast decrease $\left(\frac{dI}{dt} < 0, \frac{d^2 I}{dt^2} > 0\right)$
4. Phase IV: slow decrease $\left(\frac{dI}{dt} < 0, \frac{d^2 I}{dt^2} > 0\right)$

Estimation of parameters for each country and for $t_f$ =88 is summarized in Table 1.

Table 1 Timing of the phases of the epidemic.

| | China ($t_f$=85) | S. Korea | France | Germany | Italy | Spain | Iran | Turkey | U. Kingdom | U. States |
|---|---|---|---|---|---|---|---|---|---|---|
| $\Re_0$ | 3 | 8 | - | - | - | - | - | - | - | - |
| T | 9 | 25 | - | - | - | - | - | - | - | - |
| $t_m$ (Estimated) | 26 | 50 | 86 | 76 | 81 | 81 | 75 | 88-92 | 87-92 | 90-92 |
| $t_m$ (Real) | 27 | 50 | 84 | 76 | 89 | 93 | 75 | 93 | N/A | Not occur |
| $t_a$ | 18 | 41 | 74 | 65 | 65 | 67 | 63 | 78 | 77 | 75 |
| $t_b$ | 35 | 59 | 95-104 | 88-90 | 97-100 | 93-95 | 86-88 | 96-104 | 100-108 | 97-102 |

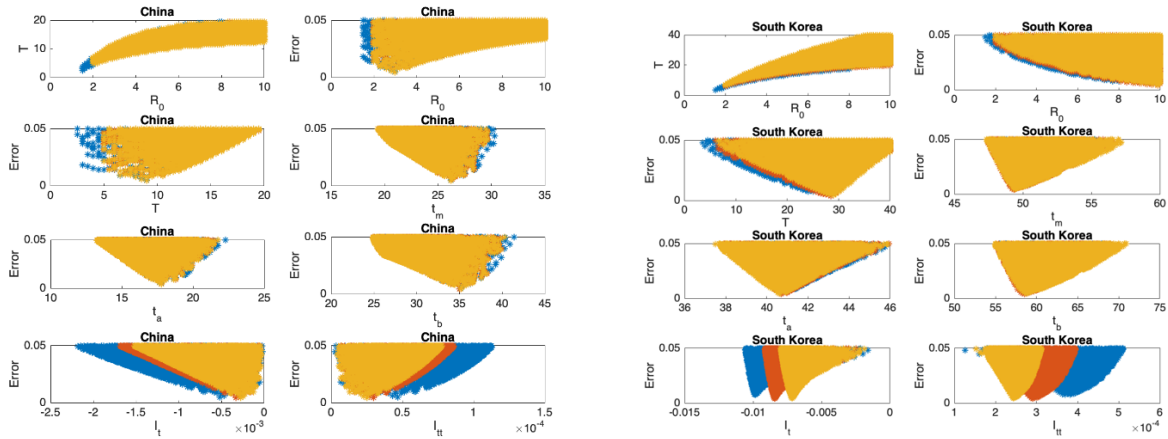

Figure 5. **China**: The 7th and 8th graphs indicate that the epidemic is in Phase IV. **South Korea**: The 7th and 8th graphs indicate that the epidemic is in Phase IV. The values for $\Re_0$ and T don't seem to fall in reasonable ranges and the data for South Korea should be studied more closely.

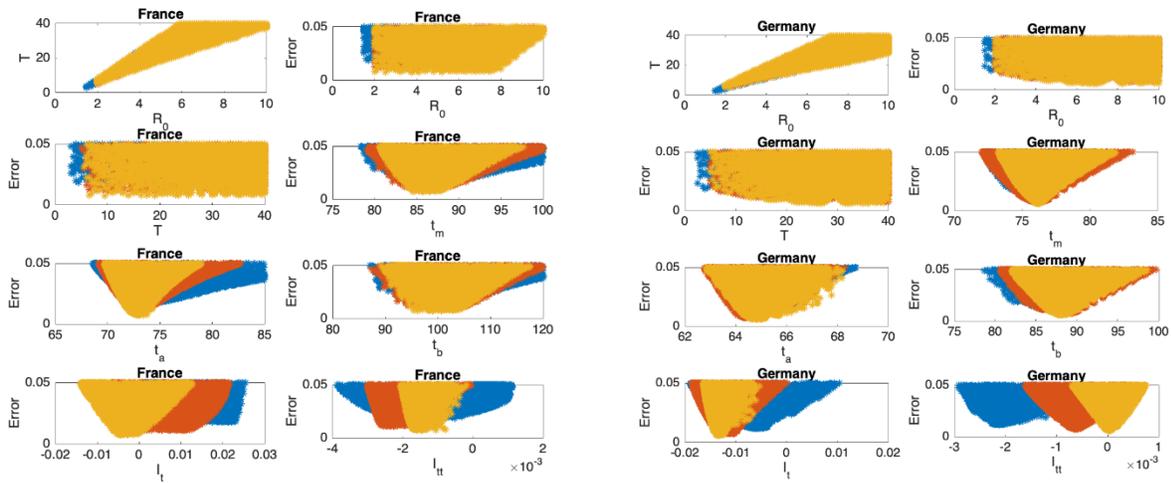

Figure 6. **France**: The 7th and 8th graphs indicate that the epidemic is at the beginning of Phase III. **Germany**: The 7th and 8th graphs indicate that the epidemic is at the beginning of Phase IV.

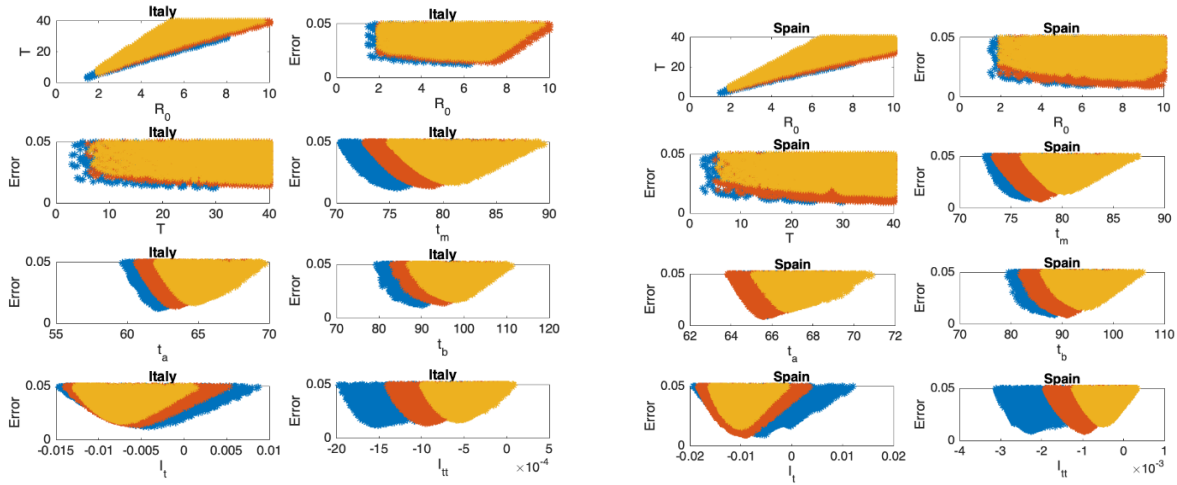

Figure 7. **Italy:** The 7th and 8th graphs indicate that the epidemic is in Phase III. **Spain**: The 7th and 8th graphs indicate that the epidemic is in Phase III.

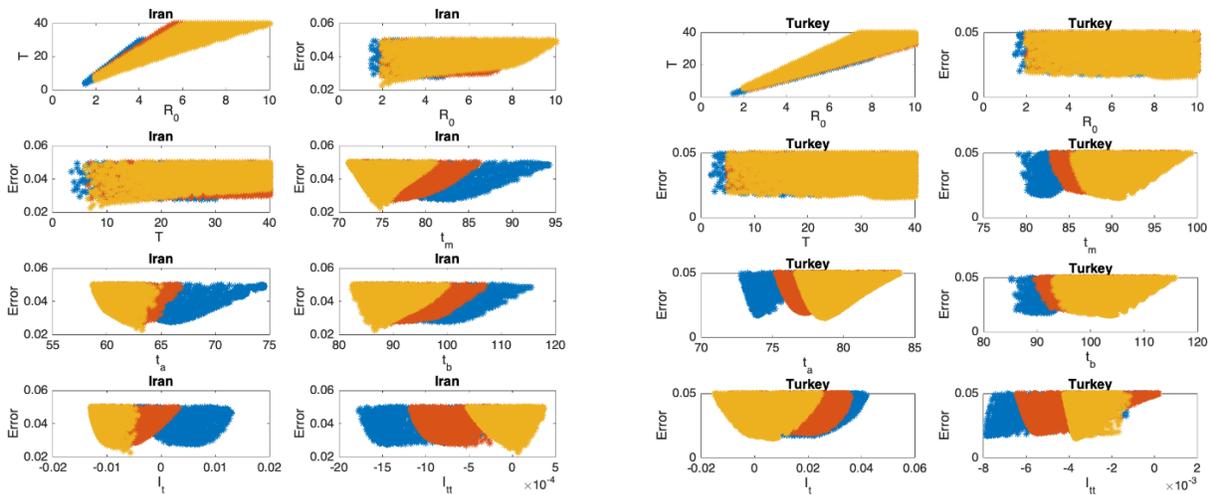

Figure 8. **Iran**: The 7th and 8th graphs indicate that the epidemic is at the beginning of Phase IV. **Turkey**: The 7th and 8th graphs indicate that the epidemic is at the end of Phase II.

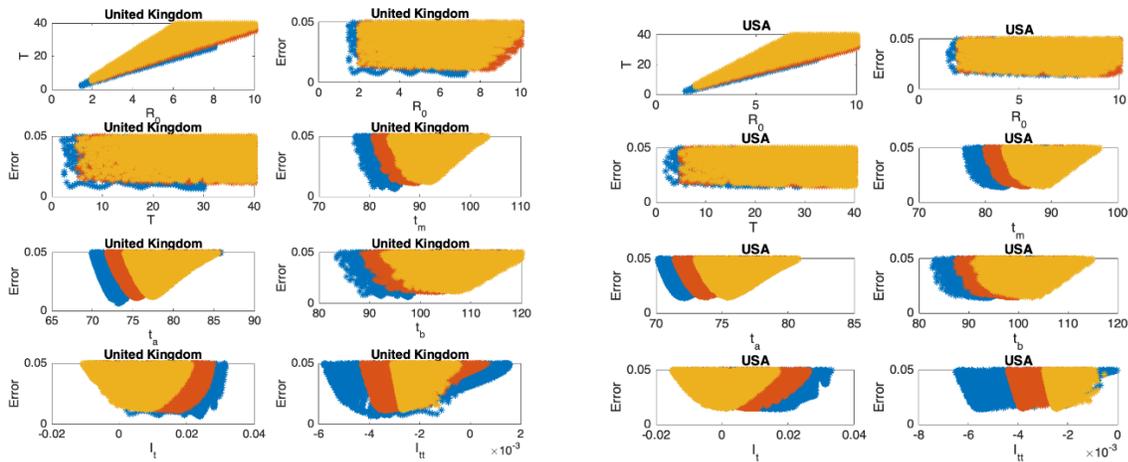

Figure 9. **United Kingdom**: The 7th and 8th graphs indicate that the epidemic is in Phase II. **United States**: The 7th and 8th graphs indicate that the epidemic is at the end of Phase II.

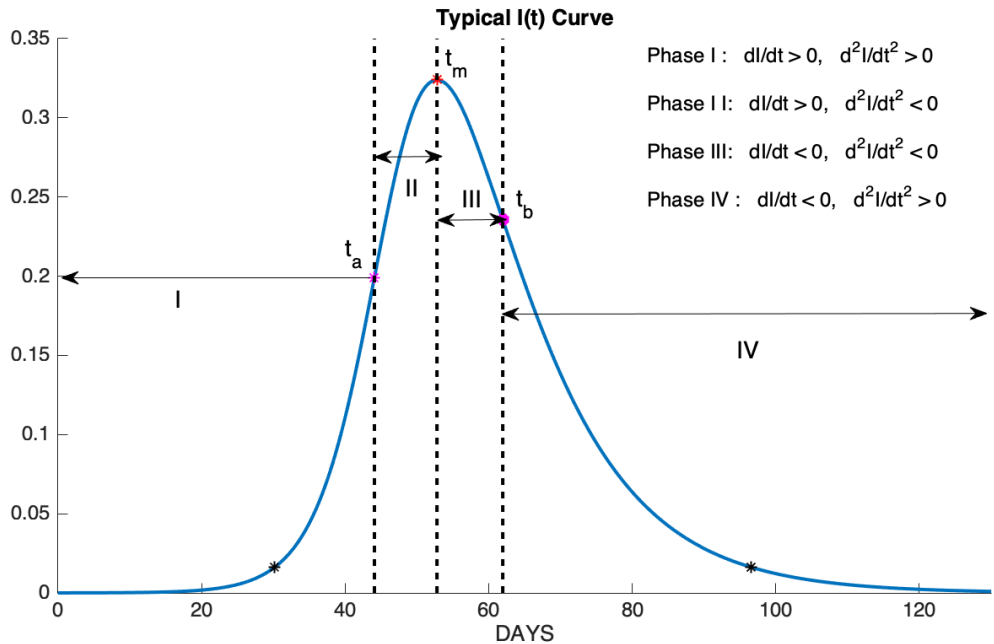

Figure 10.  Phase I: slow increase, Phase II: fast increase, Phase III: fast decrease, Phase IV: slow decrease.

## 6    MODELLING VERSUS FORECAST

In Section 5, it can be seen that although $\Re_0$ and T cannot be determined, it was possible to estimate $t_m$, $t_a$ and $t_b$ quite sharply from data.  In this section, the reliability of these estimates is discussed by comparing predictions based on data with different time spans.

The best SIR models fitting data for 78, 83 and 88 days are obtained, and data and graphs of 10 best models for each time span are plotted in Figures 11-12. For China and South Korea, for which the epidemic cycle is more or less complete, estimations based on time spans varying by 5 days give the same result as can be observed in Figure 11.

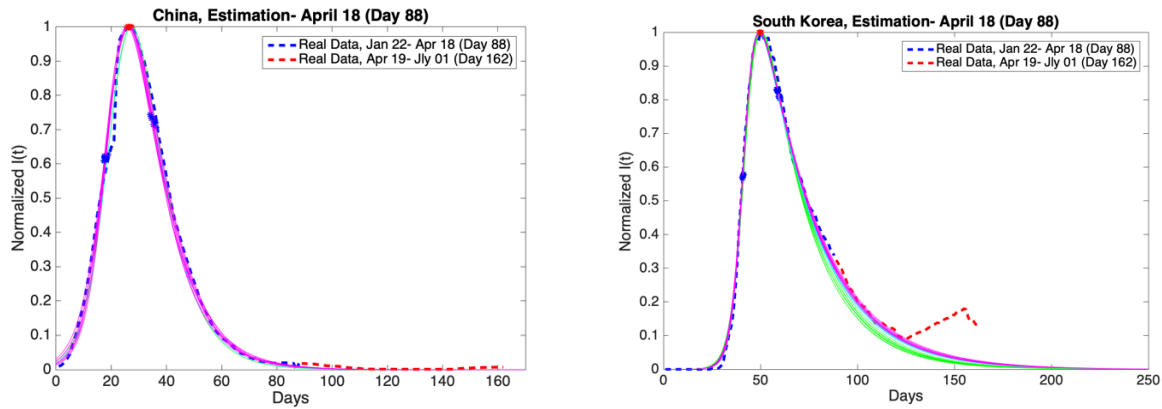

Figure 11. China and South Korea: Graphs of estimation of normalized I(t) curves for the best 10 SIR models for each time span (blue dashed curve: real data till the day 88, red dashed curve: real data between the day 88 and day 162). Accuracy of estimates was ascertained through comparison with the real data post 18th April. Data is normalized by dividing with the maximum value of infectious cases between day1 and day 88.

On the other hand, for those countries that are as yet before or around the peak of the epidemic, the situation may be different, as can be observed in Figure 12.

Accuracy of estimates was ascertained through comparison with the real data between 19th April and 1st July. These comparisons are given in Figure 11 and Figure 12 as red dashed curves. The observations are as follows.

When the initial analysis was performed, China and South Korea were in Phase 4. Our estimates and the real data for both countries are consistent.

The estimate for France is not consistent with the real data post day 88. French authorities loosened quarantine restrictions on 11th May (day 111). This event may be the reason for the fluctuations in the number of infectious cases.

The estimates for Germany and Iran are consistent with the real data. However, Germany is going through the third and the fourth phases faster than expected. Besides, the active infectious cases post 3rd May (day 103) show a continuous increase. The decrease in the active infectious cases up to this date was close to our estimates.

The estimates for Italy is consistent with the real data. On the other hand, the maximum of the infectious cases occurred slightly later than expected. In addition, Italy is going through the third and the fourth phases more slowly than expected. The most recent data conforms closely to our predictions.

Spain has not shared the data for daily discharged patients since the19th of May (day 119). Therefore, the estimates are compared with the real data up to 18th May (day 118). Our estimates and the real data for Spain are consistent. However, the maximum of the infectious cases occurred slightly later than expected. In addition, Spain is going through the third and the fourth phases more slowly than expected as in Italy.

Our estimates and the real data for Turkey are consistent. The maximum of the infectious cases occurred slightly later than expected. The decrease in the active infectious cases was close to our estimates up to a certain date. Later, the number of infectious cases shows fluctuations. Loosening quarantine restrictions on 1st June may be the reason for these fluctuations.

United Kingdom has not shared the data for daily discharged patients for a long time. We can not compare our estimation with the real data.

As for the USA the spread of the epidemic has been beyond all predictions and it is still growing.

The discrepencies between estimates and real data and the failure to estimate parameteres for USA can be explained as follows. The basic reproduction number $\Re_0$ is beta/eta and beta is a product of the virulence of the virus and the contact rate in the society. The contact rate depence crucially on lock-down measures. As these measures change, the course of the epidemic follows a different dynamic.

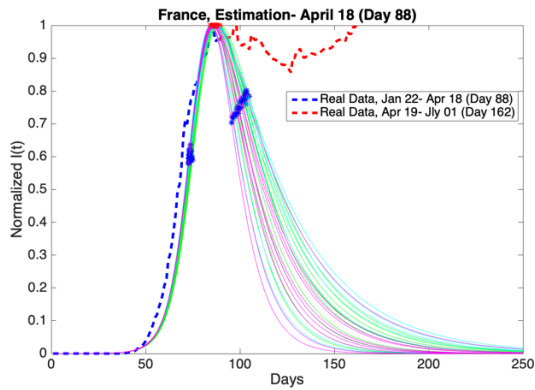
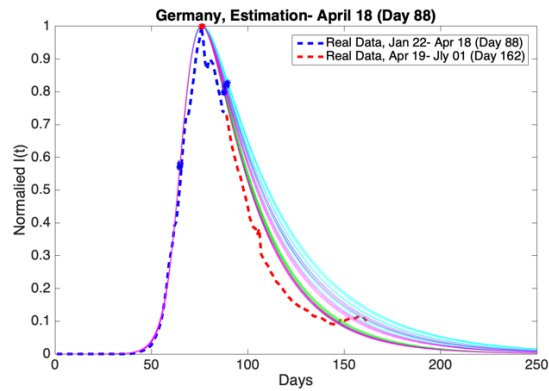

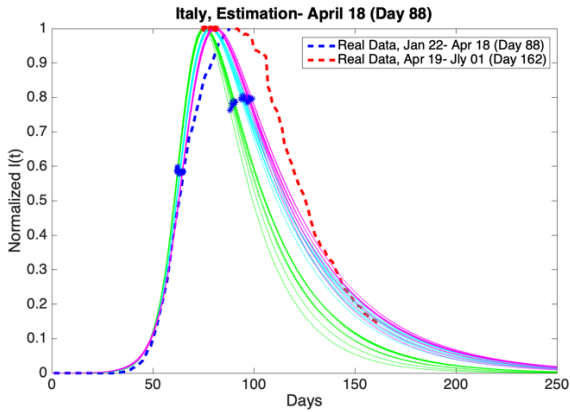
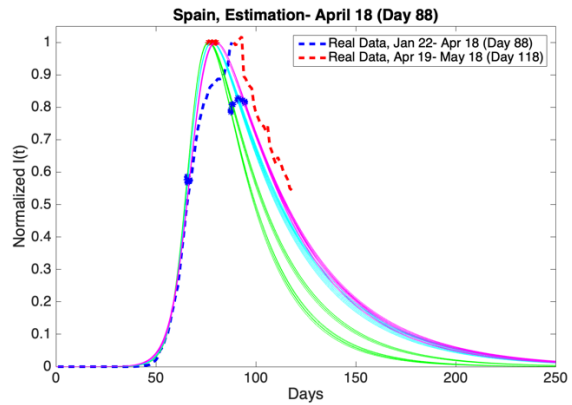

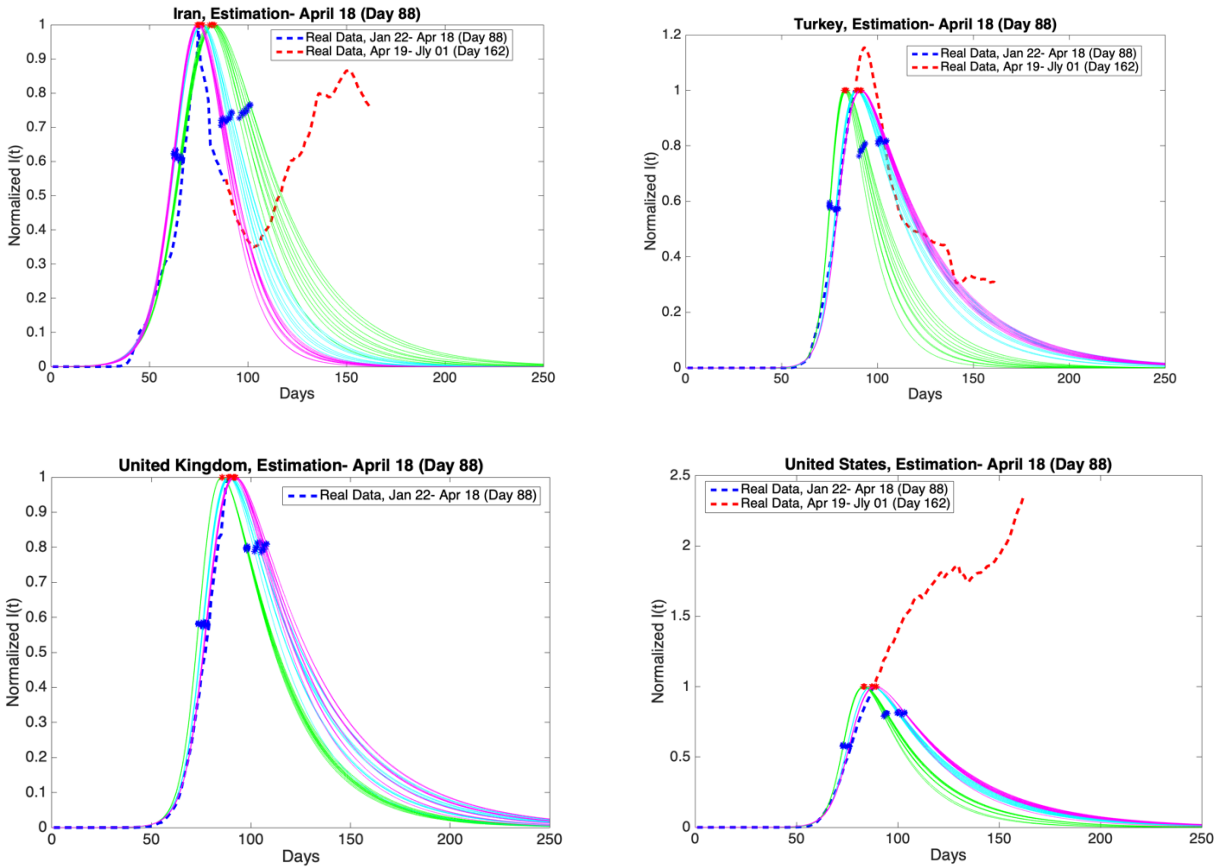

Figure 12. Graphs of estimation of normalized I(t) curves for the best 10 SIR models for each time span (blue dashed curve: real data till the day 88, red dashed curve: real data between the day 88 and day 162 except Spain and the United Kingdom) and the countries France, Germany, Italy, Spain, Iran, Turkey, the United Kingdom and the United States. Accuracy of estimates was ascertained through comparison with the real data post 18th April. Data is normalized by dividing with the maximum value of infectious cases between day1 and day 88.

## 7    CONCLUSION

The epidemic parameters of Covid-19 for ten selected countries are estimated by using the data released by the state offices. These parameters include the basic reproduction number, mean duration of infectious period, the time at which the number of infectious cases reaches its maximum, the time at which the rate of increase in the number of infectious cases reaches its maximum, the time at which the rate of decrease in the number of infectious cases reaches its maximum. For each country, the best Susceptible-Infected-Removed (SIR) models fitting *cumulative case data* are obtained. A wide variety of intervals with different scales of the parameters, basic reproduction number $\Re_0$ and infectious period T, are observed. More specifically, the basic reproduction number and mean duration of infectious period are estimated only for China since the spread of the disease there is over. These parameters are found to be 3 and 5, respectively. The fact that the median incubation and infection periods are approximately 5 days, supports the observations for $\Re_0$ and T. However, the basic reproduction number and infectious period for other countries cannot be predicted from the normalized data but the timing of key events can be estimated quite reliably. To summarize, we show that the quantity that can be the most robustly estimated from the normalized data, is the timing of the highest rate of increase in the number of infections, i.e, the inflection point of the number of infected individuals. However, it should be pointed out that the analysis performed by the SIR model for South Korea provides dissimilar results which can be explained by the unique age distribution nature of the confirmed cases.

## 8    CONFLICT OF INTEREST

The authors declare that the research was conducted in the absence of any commercial or financial relationships that could be construed as a potential conflict of interest.

## 9    AUTHOR CONTRIBUTIONS

AHB and AD performed computations; AD collected the data; OE provided medical insights and SA and APD performed literature survey and wrote the paper.